# Observation of the magnon Hall effect

**[One-sentence summary]**
We have observed the Hall effect of magnons for the first time in terms of transverse thermal transport in a ferromagnetic insulator $Lu_2V_2O_7$ with pyrochlore structure.

Y. Onose,[1,2,†] T. Ideue,[1] H. Katsura,[3] Y. Shiomi,[1,4] N. Nagaosa,[1,4] and Y. Tokura[1,2,4]

[1] Department of Applied Physics, University of Tokyo, Tokyo 113-8656, Japan

[2] Multiferroics Project, ERATO, Japan Science and Technology Agency (JST), c/o Department of Applied Physics, University of Tokyo, Tokyo 113-8656, Japan

[3] Kavli Institute for Theoretical Physics, University of California, Santa Barbara, CA 93106, USA

[4] Cross-Correlated Materials Research Group (CMRG) and Correlated Electron Research Group (CERG), Advanced Science Institute, RIKEN, Wako, 351-0198, Japan

† To whom correspondence should be addressed E-mail: onose@ap.t.u-tokyo.ac.jp

**[abstract]**

The Hall effect usually occurs when the Lorentz force acts on a charge current in a conductor in the presence of perpendicular magnetic field. On the other hand, neutral quasi-particles such as phonons and spins can carry heat current and potentially show the Hall effect without resorting to the Lorentz force. We report experimental evidence for the anomalous thermal Hall effect caused by spin excitations (magnons) in an insulating ferromagnet with a pyrochlore lattice structure. Our theoretical analysis indicates that the propagation of the spin wave is influenced by the Dzyaloshinskii-Moriya spin-orbit interaction, which plays the role of the vector potential as in the intrinsic anomalous Hall effect in metallic ferromagnets.

**[text]**

Electronics based on the spin degree of freedom, (spintronics), holds promise for new developments beyond Si-based-technologies (*1*), and avoids the dissipation from Joule heating by replacing charge currents with currents of the magnetic moment (spin currents). Phenomenon such as the spin Hall effect (generation of a transverse spin current by a longitudinal electric field) in metals and semiconductors have therefore attracted much attention recently (*2*). However, some dissipation is still inevitable because the spin current in these conducting materials is carried by electronic carriers. In this sense, spin transport in insulating magnets seems more promising.

In magnetic insulators, the spin moments are carried by magnons, which are quanta of magnetic excitations. A fundamental question for the magnon spin current is whether it exhibits the Hall effect. The Hall effect is usually driven by the Lorentz force, and therefore uncharged particles (photons, phonons, magnons, etc.) may be expected not to show it. In ferromagnets, however, the Hall effect proportional to the magnetization, termed anomalous Hall effect, can be driven by the relativistic spin-orbit interaction, and does not require the Lorentz force (*3*). Moreover, the Hall effect of photons (*4,5,6*) and that of phonons (*7,8,9*) have been already predicted and experimentally observed. However, the Hall effect of magnons, that would be most relevant to spin-current electronics, has been challenging to observe.

We report the magnetic and thermal-transport properties of an insulating collinear ferromagnet $Lu_2V_2O_7$ with pyrochlore structure (Fig. 1A). Figure 1A shows the vanadium sublattice in $Lu_2V_2O_7$ with a pyrochlore structure, which is composed of corner sharing tetrahedra. This structure can be viewed as a stacking of alternating Kagomé and triangular lattices along the [111] direction. The spin polarized neutron diffraction suggested the orbitals of the *d* electron are ordered so that all the orbitals point to the center of mass of the V tetrahedron (*10*). A theoretical calculation shows that the virtual hopping process to the high-energy state stabilizes the ferromagnetic order of V spin in this orbital ordered state (*10*). In the pyrochlore structure, because the midpoint between the two apices of a tetrahedron is not an inversion symmetry center, there is the Dzyaloshinsky-Moriya (DM) interaction $H_{DM} = \sum_{<ij>} \vec{D}_{ij} \cdot (\vec{S}_i \times \vec{S}_j)$, where $\vec{D}_{ij}$, $\vec{S}_i$ are, respectively, DM vector between *i* and *j* sites and V spin moment at *i* site. As shown in Fig. 1B, the DM vector $\vec{D}_{ij}$ is perpendicular to the vanadium bond and parallel to the surface of the cube indicated by gray lines, according to the crystal symmetry. Even in the perfectly collinear ferromagnetic ground state spin configuration, therefore, the DM interaction affects the spin wave and gives rise to the thermal Hall effect as discussed below.

Figure 2 illustrates the magnetic, electric, and thermal properties of $Lu_2V_2O_7$. The spontaneous magnetization *M* emerges below $T_C$=70 K (Fig. 2A). Figure 2B shows the magnetization curves along various magnetic field directions at 5 K. The magnetization saturates at relatively low field (less than 1 T), and the saturated magnetization is isotropic and almost coincides with 1 $\mu_B$, indicating the collinear *S*=1/2 ferromagnetic state above the saturation field. The resistivity increases rapidly with decreasing temperature *T* (Fig. 2C). The longitudinal thermal conductivity (*11*) monotonically falls with decreasing temperature (Fig. 2D). The magnitude is small compared with usual insulators, but comparable to similar orbital ordered materials (*12*). According to the Wiedemann-Franz law, the electric contribution of thermal conductivity is less than $10^{-5}$ W/Km below 100 K. Therefore, the heat current is carried only by phonons and magnons in this temperature region. By the analysis of the magnetic field variation of the thermal conductivity (*13*), we estimate the mean free paths of phonon and magnon ($l_{ph}$ and $l_{mag}$) at 20 K as $l_{ph}$=3.5 nm, and $l_{mag}$=3.6 nm. Because the obtained $l_{ph}$ and $l_{mag}$ are much larger than the V-V distance

(=0.35 nm), the Bloch waves of phonons and magnons are well defined at least at 20 K.

The thermal Hall effect (the Righi-Leduc effect) is usually induced by the deflection of electronic heat current by the magnetic field in metallic materials. As the heat current in this material is carried only by phonons and magnons and not by charge-carriers, any observed thermal Hall conductivity would provide evidence for the Hall effect of phonons and/or magnons. Our measurements of thermal Hall conductivity are presented in Fig. 3. Below $T_C$=70 K, the signal is well restored, whereas it is quite small above 80 K. The magnitude of the thermal Hall conductivity has a maximum at around 50 K. Similar to the magnetization, the thermal Hall conductivity steeply increases and saturates in the low-magnetic field region. Thus, what is presently observed in the heat transport is not the normal Hall effect proportional to the magnetic field strength, but the anomalous (spontaneous) Hall effect affected by the spontaneous magnetization. However, the thermal Hall conductivity gradually decreases with magnetic field after saturation in the low temperature region; this can be explained by the magnon gap induced by the magnetic field as discussed below. Figure 4A shows the temperature dependence of the spontaneous thermal Hall conductivity (the thermal Hall conductivity just above the saturation field) for $H$||[100], [110], and [111]; it is independent of the field direction within the error bars.

The thermal Hall effect caused by phonons has been reported in $Tb_3Ga_5O_{12}$ (*7*) and explained by the spin-phonon interaction (*8,9*). The intrinsic mechanism in (*8*) does not depend on the magnon population. The mean free path of phonons is expected to increase with magnetic field as a consequence of reduced scattering by magnetic fluctuations. Therefore, our observation of the decrease of the thermal Hall conductivity in the high field region cannot be explained in terms of the phonon mechanism. On the other hand, the reduction of the magnon population, as reflected by the behavior of the specific heat under the magnetic field (*13*), will diminish the magnon contribution of the thermal Hall conductivity. In the mechanism based on the scattering of phonons by spins (*9*), the thermal Hall angle $\kappa_{xy}/\kappa_{xx}$ is anticipated to be proportional to the magnetization such as in the case of $Tb_3Ga_5O_{12}$ (*7*). For $Lu_2V_2O_7$, the $\kappa_{xy}$ steeply decreases with increasing temperature around $T_C$ even faster than the magnetization as shown in Fig. 4B, which contradicts the scenario of phonon Hall effect. On the other hand, magnons propagate in terms of the exchange interaction (see Eq. 3 below), and therefore cannot be valid in the magnetic-field induced spin-polarized state above $T_C$. Hence, the temperature dependence can be well explained in terms of the magnon Hall effect.

We now turn to the quantitative calculation of the thermal Hall effect in $Lu_2V_2O_7$. This material is a ferromagnetic Mott-insulator and the spin-1/2 $V^{4+}$ ions form a pyrochlore lattice (Fig. 1A). The effective spin Hamiltonian describing this system is given by

$$H_{eff} = \sum_{\langle ij \rangle} -J\vec{S}_i \cdot \vec{S}_j + \vec{D}_{ij} \cdot (\vec{S}_i \times \vec{S}_j) - g\mu_B \vec{H} \cdot \sum_i \vec{S}_i, \qquad (1)$$

where $-J$ is the nearest neighbor ferromagnetic exchange and the last term is the Zeeman coupling with an external field $\vec{H}$. Note that the DM interaction does not disturb the perfect ferromagnetic alignment of the spins along the direction of a weak external magnetic field $\vec{H}$ because the sum of the DM vectors on the bonds sharing the same site is zero. Because we are interested in the low temperature regime $T \ll T_C$, the interactions between magnons are neglected. Therefore, we consider the Bloch state of a single magnon,

$$|\vec{k}\rangle = \frac{1}{\sqrt{N}} \sum_i e^{i\vec{k} \cdot \vec{R}_i} |i\rangle, \qquad (2)$$

where $|i\rangle$ is the state representing the spin at site $i$ being along the $-\vec{H}$ direction with all the other spins pointing in $\vec{H}$ direction. The matrix element corresponding to the transfer of magnons reads as

$$\langle i| -J\vec{S}_i \cdot \vec{S}_j + \vec{D}_{ij} \cdot (\vec{S}_i \times \vec{S}_j) |j\rangle$$

$$= \langle i| -\frac{J}{2}(S_i^+ S_j^- + S_i^- S_j^+) + \frac{iD_{ij}}{2}(S_i^+ S_j^- - S_i^- S_j^+) |j\rangle = -\frac{\tilde{J}}{2} e^{i\phi_{ij}} \qquad (3)$$

where $S^\pm$ is the operator which increases or decreases the spin component along the direction $\vec{n} = (n_x, n_y, n_z) = \vec{H}/|\vec{H}|$, $D_{ij} = \vec{D}_{ij} \cdot \vec{n}$, and $\tilde{J}e^{i\phi_{ij}} = J + iD_{ij}$. Thus the DM interaction acts as a vector potential $\phi_{ij}$ and serves as the "orbital magnetic field" for the propagation of magnons. The spin-wave Hamiltonian becomes just a tight-binding Hamiltonian on a pyrochlore lattice with phase factors. Thanks to the different types of loops in the unit cell of the pyrochlore lattice, the orbital magnetic field avoids the cancellation and gives rise to the Hall effect (*14*).

The formula for the thermal Hall conductivity $\kappa_{\alpha\beta}$ of the magnon system was derived in (*14*). In the low-temperature region, the dominant contribution to $\kappa_{\alpha\beta}$ comes from the lowest magnon band and small $k$ due to the Bose distribution function. Retaining only the first-order terms in the DM interaction, the analytic expression for the anomalous thermal Hall conductivity due to magnons is obtained as (*13*)

$$\kappa_{\alpha\beta}(H,T) = \Phi_{\alpha\beta} \frac{k_B^2 T}{\pi^{3/2} \hbar a} \left(2 + \frac{g\mu_B H}{2JS}\right)^2 \sqrt{\frac{k_B T}{2JS}} \text{Li}_{5/2}\left[\exp\left(-\frac{g\mu_B H}{k_B T}\right)\right], \quad (4)$$

with the polylogarithm $\text{Li}_n(z)$ being given by $\text{Li}_n(z) = \sum_{k=1}^{\infty} \frac{z^k}{k^n}$.

Here, $a$ is the lattice constant, and $\Phi_{\alpha\beta} = -\varepsilon_{\alpha\beta\gamma} n_\gamma D/(8\sqrt{2}J)$ with the totally antisymmetric tensor $\varepsilon_{\alpha\beta\gamma}$ and $D = |\vec{D}_{ij}|$. In Fig. 4C, we show the fitting of the data at $T$=20K to this expression with $JS=8D_S/a^2$, where $D_S$ is the spin stiffness constant obtained from the analysis of specific heat shown in the supporting online material (*13*). Therefore, the only parameter is the ratio of DM interaction strength $D$ to the exchange coupling $J$. The agreement between experiment and theory is obtained for $D/J$ = 0.32; this is a reasonable value for transition-metal oxides. In the spinel-type antiferromagnet $CdCr_2O_4$, for example, $D/J$ has been estimated as 0.19 from the measured pitch of the spiral spin structure and ab initio calculations (*15*), which is of the same order of magnitude as our result.

We have observed the magnon Hall effect in the ferromagnetic insulator $Lu_2V_2O_7$. The magnon spin current in insulating materials has been discussed rarely (*16*) other than the spontaneous spin current relevant to electronic polarization (*17*). The finding of the Hall effect of the magnon spin current may open a new field of spin transport in magnetic insulators.

19. The authors thank C. Terakura for the design of Fig. 1C, and P. A. Lee, S. Fujimoto and T. Arima for fruitful discussion. This work is supported by Grant-in-Aid for Scientific Research (Grants No. 17071007, 17071005, 19048008, 19048015, 19684011, 20046004, 20340086, and 21244053) from the Ministry of Education, Culture, Sports, Science and Technology of Japan, and by the Funding Program for World-Leading Innovative R&D on Science and Technology (FIRST), Japan.

**Figures**

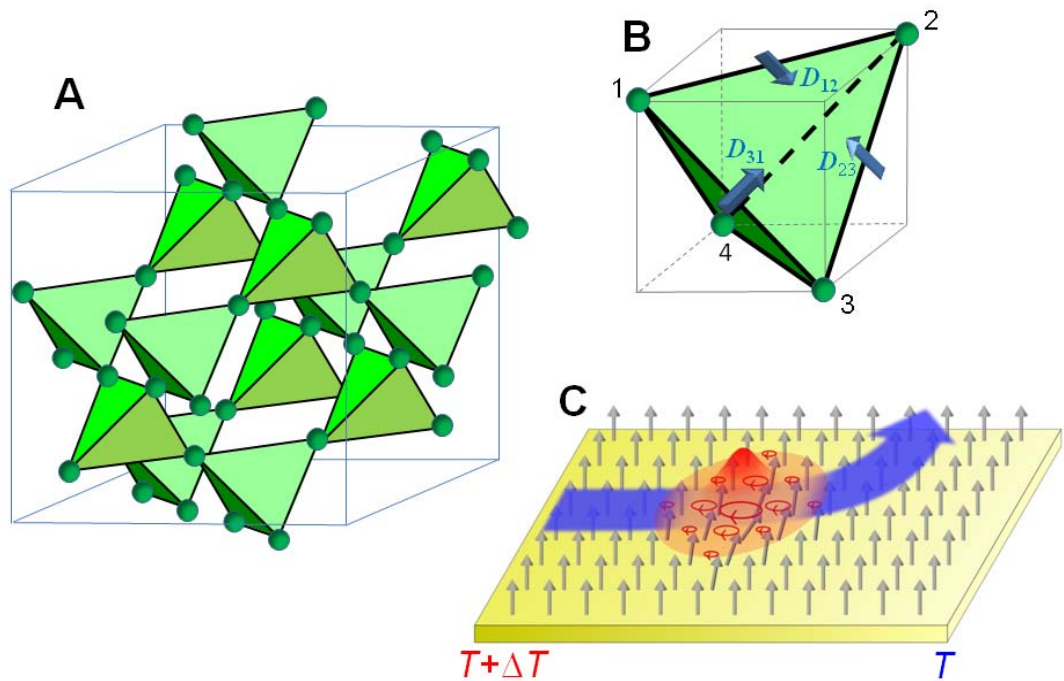

**Fig. 1.** The crystal structure of $Lu_2V_2O_7$ and the magnon Hall effect. **(A)** The V sublattice of $Lu_2V_2O_7$, composed of corner-sharing tetrahedra. **(B)** The direction of the Dzyaloshinskii-Moriya vector $\vec{D}_{ij}$ on each bond of the tetrahedron. The Dzyaloshinskii-Moriya interaction $\vec{D}_{ij} \cdot (\vec{S}_i \times \vec{S}_j)$ acts between the *i* and *j* sites. **(C)** The magnon Hall effect: A wave packet of magnon (a quantum of spin precession) moving from the hot to the cold side is deflected by the Dzyaloshinskii-Moriya interaction playing the role of a vector potential.

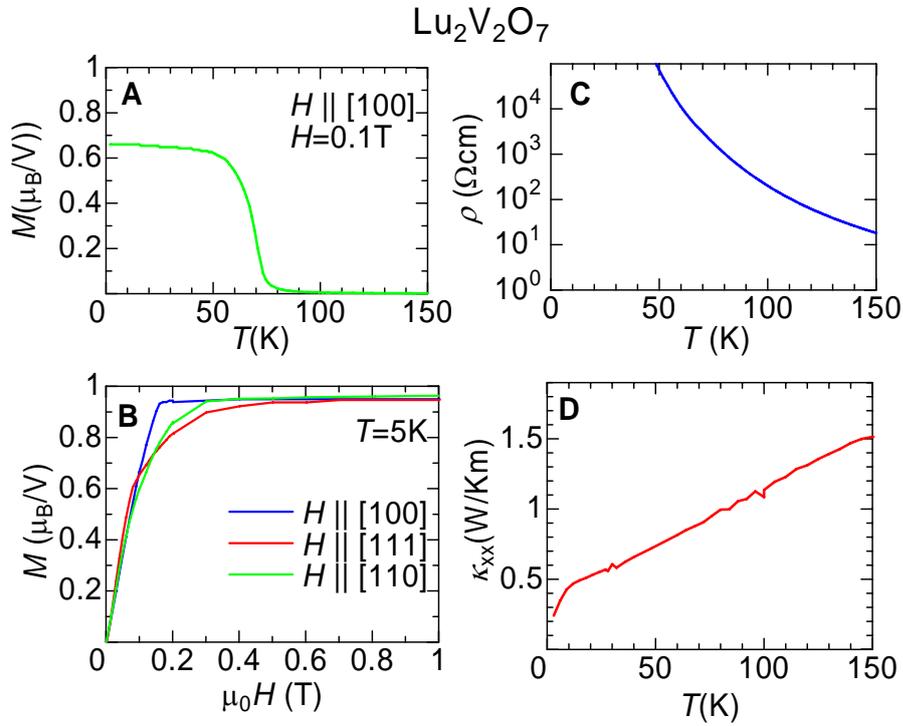

**Fig. 2.** Magnetic, electric, and thermal properties of Lu$_2$V$_2$O$_7$. **(A)** Temperature dependences of magnetization at the magnetic field $H$=0.1T along the [100] direction. **(B)** Magnetization curves at $T$=5 K for $H$||[100], $H$||[110], and $H$||[111]. **(C)** Temperature variation of resistivity $\rho$. **(D)** Temperature variation of longitudinal thermal conductivity $\kappa_{xx}$.

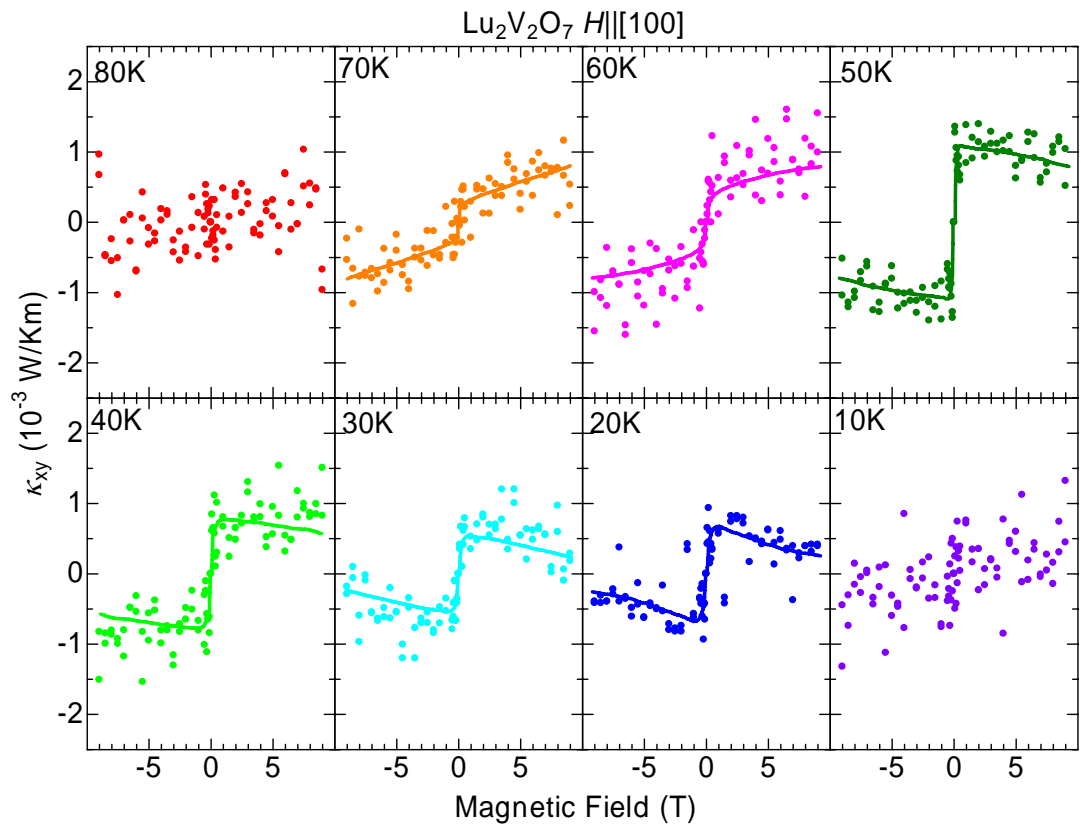

**Fig. 3.** Magnetic field variation of the thermal Hall conductivity of $Lu_2V_2O_7$ at various temperatures. The magnetic field is applied along the [100] direction. The solid lines are guides to the eye.

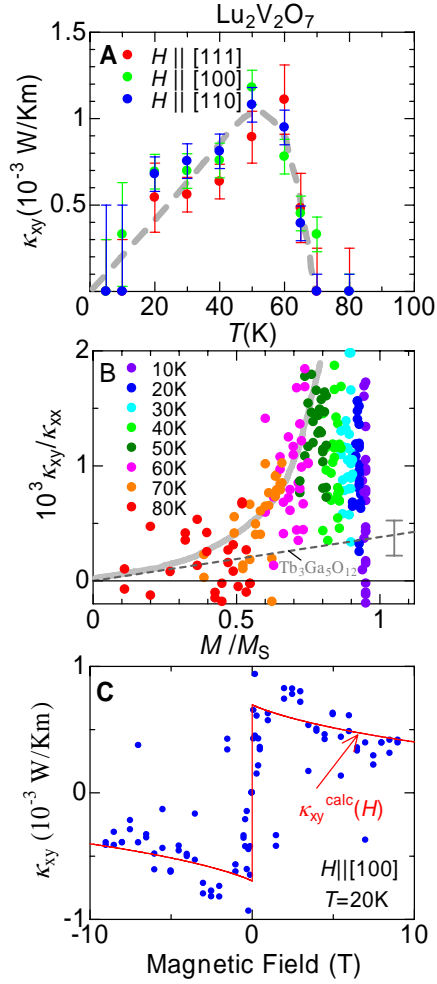

**Fig. 4.** **(A)** Temperature dependence of the spontaneous thermal Hall conductivity (the thermal Hall conductivity just above the saturation field) for $H\|[100]$, $H\|[110]$, and $H\|[111]$. The thick dashed line is a guide to the eye. **(B)** The thermal Hall angle $\kappa_{xy}/\kappa_{xx}$ plotted against the magnetization ($M$). For $Tb_3Ga_5O_{12}$ (dashed line), the value of $\kappa_{xy}/\kappa_{xx}$ divided by the magnetic field $H$ is taken from (7), and the magnetic susceptibility ($M/H$) is estimated from the magnetization curves in (18). The thick solid line is a guide to the eye. **(C)** Magnetic field variation of the thermal Hall conductivity at 20 K for $H\|[100]$. The red solid line indicates the magnetic field dependence given by the theory (Eq. 4) based on the Dzyaloshinskii-Moriya interaction.